\documentclass[conference]{IEEEtran}
\IEEEoverridecommandlockouts
\usepackage{cite}
\usepackage{amsmath,amssymb,amsfonts}
\usepackage{graphicx}
\usepackage{textcomp}
\usepackage{xcolor}
\usepackage{multirow}
\usepackage{kotex}
\usepackage{multicol,multirow}
\usepackage{makecell}
\usepackage{booktabs}

\usepackage[super]{nth}
\def\BibTeX{{\rm B\kern-.05em{\sc i\kern-.025em b}\kern-.08em
    T\kern-.1667em\lower.7ex\hbox{E}\kern-.125emX}}
\begin{document}

\title{Calibration-Free Driver Drowsiness Classification based on Manifold-Level Augmentation 
\footnote{{\thanks{20xx IEEE. Personal use of this material is permitted. Permission
from IEEE must be obtained for all other uses, in any current or future media, including reprinting/republishing this material for advertising or promotional purposes, creating new collective works, for resale or redistribution to servers or lists, or reuse of any copyrighted component of this work in other works. This work was partly supported by Institute of Information and Communications Technology Planning and Evaluation (IITP) grants funded by the Korea government (No. 2017-0-00451, Development of BCI based Brain and Cognitive Computing Technology for Recognizing User’s Intentions using Deep Learning; No. 2019-0-00079, Artiﬁcial Intelligence Graduate School Program (Korea University); No. 2021-0-00866, Development of BMI application technology based on multiple bio-signals for autonomous vehicle drivers). }
}}}

\author{\IEEEauthorblockN{Dong-Young Kim}
\IEEEauthorblockA{\textit{Dept. Artificial Intelligence} \\
\textit{Korea University}\\
Seoul, Republic of Korea \\ 
dy\_kim@korea.ac.kr}
\and
\IEEEauthorblockN{Dong-Kyun Han}
\IEEEauthorblockA{\textit{Dept. Brain and Cognitive Engineering} \\
\textit{Korea University}\\
Seoul, Republic of Korea \\
dk\_han@korea.ac.kr}
\and
\IEEEauthorblockN{Hye-Bin Shin}
\IEEEauthorblockA{\textit{Dept. Brain and Cognitive Engineering} \\
\textit{Korea University}\\
Seoul, Republic of Korea \\
hb\_shin@korea.ac.kr}
}

\maketitle

\begin{abstract}
Drowsiness reduces concentration and increases response time, which causes fatal road accidents. Monitoring drivers' drowsiness levels by electroencephalogram (EEG) and taking action may prevent road accidents. EEG signals effectively monitor the driver's mental state as they can monitor brain dynamics. However, calibration is required in advance because EEG signals vary between and within subjects. Because of the inconvenience, calibration has reduced the accessibility of the brain-computer interface (BCI). Developing a generalized classification model is similar to domain generalization, which overcomes the domain shift problem. Especially data augmentation is frequently used. This paper proposes a calibration-free framework for driver drowsiness state classification using manifold-level augmentation. This framework increases the diversity of source domains by utilizing features. We experimented with various augmentation methods to improve the generalization performance. Based on the results of the experiments, we found that deeper models with smaller kernel sizes improved generalizability. In addition, applying an augmentation at the manifold-level resulted in an outstanding improvement. The framework demonstrated the capability for calibration-free BCI. 
\end{abstract}

\begin{small}
\end{small}

\section{INTRODUCTION}
Drowsy driving continues to cause accidents. Therefore, an accurate drowsiness state classification is necessary to prevent road traffic accidents. Physiological signals are frequently used to estimate mental states. Especially, EEG is capable of measuring brain activity directly and monitoring brain activity \cite{won2017motion, gao2019eeg, lee2019connectivity}. Recently, detecting affective states such as emotions and mental states with EEG signals, in other words, affective brain-computer interface (BCI), has consistently gained interest \cite{wu2020transfer, thung2018conversion, zhang2017hybrid}. For instance, Xu \textit{et al.} \cite{xu2021key} proposed a unified convolutional attention neural network that concurrently identifies personal information and detects the driver's drowsiness. Paulo \textit{et al.} \cite{paulo2021cross} proposed a drowsiness detection model with spatiotemporal image encoding representations such as recurrence plots and gramian angular fields.

EEG signals, however, have a characteristic of variability which differs between and within subjects \cite{kwon2019subject,cui2019eeg}. Though it is common to have different signals between subjects, different characteristics of EEG signals are also obtained according to time, physiological state, and the location of measurement. Usually, calibration is required to readjust the system to the current users' state \cite{kim2019subject, lee2018high}. This user-unfriendly repetition has been an obstacle to practical BCI \cite{cui2019eeg, suk2014predicting}. 

One approach to overcome the variability problem and develop a generalized drowsiness classification framework without calibration is transfer learning \cite{kostas2020thinker}. For instance, Liu \textit{et al.} \cite{liu2020inter} extracted power spectral density features with transfer learning-based algorithms, transfer component analysis, and maximum independence domain adaptation. In particular, a framework that does not require subject-specific data, or a subject-independent framework, intends to generalize well on unseen subjects with only the multiple source subject data. This goal is similar to domain generalization (DG) \cite{han2021domain}. As we regard developing a subject-independent framework from a DG perspective, the variability problem can be viewed as a domain shift problem, where each subject and session can be considered an independent domain \cite{wu2020transfer}. Cui \textit{et al.} \cite{cui2019eeg} applied a DG method, episodic training, and Hwang \textit{et al.} \cite{hwang2021mitigating} applied a domain adaptation method, domain adversarial training, for a subject-independent driver drowsiness classification.

Generally, when a distributional difference between source and target data is expected, DG is a preferred task \cite{gulrajani2021in}. In computer vision literature, DG tasks are solved through perturbing source domains with data augmentation, applying domain adversarial training \cite{gulrajani2021in}, or meta-learning \cite{dou2019domain}. Among methods, data augmentation is frequently used. This method generates data samples by flipping, cropping, and sliding windows, increasing the diversity of source domains \cite{zhou2022domain}. The model can experience various data to simulate the real-world domain shift \cite{zhou2022domain}. In most cases, data augmentation is conducted with raw data samples. However, raw-level augmentation methods can potentially deteriorate the characteristics of signals. Therefore, we compare manifold-level augmentation methods and find the suitable method for driver drowsiness classification.

This paper proposes a generalized driver drowsiness classification framework without the need for subject-specific data. Each subject is considered an independent domain, and intermediate features are used for augmentation. We conducted experiments with multiple augmentation methods and claimed that augmentation in the manifold-level improves the generalization performance in the EEG signal dataset.

\section{METHODS}
We compared four deep learning models known to have shown competitive performance in EEG-based classification. We selected the highest-performing model as the backbone model for experimenting with various augmentation methods.

\subsection{Deep Learning Models}
We briefly introduce four deep learning-based models that were used for comparison: EEGNet4,2 \cite{lawhern2018eegnet}, EEGNet8,2 \cite{lawhern2018eegnet}, one-dimensional ResNet with 8 layers (ResNet1D-8) \cite{cheng2020subject, han2021domain}, and 18 layers (ResNet1D-18) \cite{cheng2020subject, han2021domain}. In EEG-based classification, EEGNet4,2 and EEBNet8,2 are commonly used and have shown competitive performance \cite{jeong2020brain, jeong2020decoding}. As with ResNet1D, a modified version of ResNet \cite{he2016deep}, has recently shown remarkable performance in DG and mental state classification \cite{han2021domain, kim2022dg}.

\subsubsection{EEGNet4,2}
EEGNet4,2 learns in a total of four temporal filters and two spatial filters per temporal filter. The model comprises temporal, spatial, depthwise, and separable convolution with kernel sizes based on the sampling rate and channel size. Depthwise and separable convolution were used for the reduction of parameters. Moreover, it has a batch normalization, average pooling, and dropout layer with an exponential linear unit (ELU) activation function. 

\subsubsection{EEGNet8,2}
This model has the same structure as EEGNet4,2, except it learns eight temporal filters and two spatial filters per temporal filter.

\subsubsection{ResNet1D-8}
ResNet1D-8 consists of three residual blocks and a fully-connected layer. Features are extracted from the residual blocks and classified with a fully-connected layer. Dropout, batch normalization, ELU activation, and one-dimensional convolutional layers are included in each residual block. The kernel size of each residual block is 11, 9, and 7, respectively. In addition, a skip connection is implemented in the residual block when there is a discrepancy between the input and output size of the residual block. The convolutional layer in the skip connection has a kernel size of 1 to match the shape. 

\subsubsection{ResNet1D-18}
ResNet1D-18 consists of four residual blocks for feature extraction and a fully-connected layer for classification. As ResNet1D-8, the residual block in this model has the same composition. Instead, since this model is deeper than ResNet1D-8, the kernel size of each convolution layer is 3 to reduce the model parameters.

\subsection{Data Augmentation Methods}
We applied three data augmentation methods: Mixup \cite{zhang2018mixup}, Manifold Mixup\cite{verma2019manifold}, and MixStyle \cite{zhou2021domain, kim2022dg}. Mixup is a raw-level augmentation method, and Manifold Mixup and MixStyle are manifold-level augmentation methods.

\subsubsection{Mixup}
Mixup performs a weighted sum of randomly selected raw input values and one-hot encoding labels from the dataset, respectively, with a weight sampled from a predefined distribution. Since Mixup sums two data, the predefined distribution is the Beta distribution. When the $i$-th data and labels are denoted as ($\textbf{x}_i$, $y_i$), the constructed virtual training sample ($\textbf{x}^{mix}$, $y^{mix}$) is as follows:
\begin{equation}
    \begin{split}
        &\textbf{x}^{mix}=\lambda \textbf{x}_i + (1-\lambda)\textbf{x}_j, \\
        &y^{mix}=\lambda y_i + (1-\lambda)y_j,
    \end{split}
    \label{originmixup}
\end{equation}
where $\lambda$ is a weight sampled from the Beta distribution ($\lambda \sim Beta(\alpha,\alpha), \alpha\in(0,\infty)$). 

\subsubsection{Manifold Mixup}
Manifold Mixup is an extended version of Mixup, which constructs virtual training samples at the manifold-level. The original raw input values are changed into the features extracted from each layer. For each pair of raw input values and class labels ($\textbf{x}_{l,i}$, $y_{l,i}$), let $\textbf{z}_{l,i}$ be the extracted feature of $x_{l,i}$ after the $l$-th residual block. The Eq. \ref{originmixup} is modified as 
\begin{equation}
    \begin{split}
        &\textbf{z}^{mix}_l=\lambda \textbf{z}_{l,i} + (1-\lambda)\textbf{z}_{l,j}, \\
        &y^{mix}_l=\lambda y_{l,i} + (1-\lambda)y_{l,j}.
    \end{split}
\end{equation}

\subsubsection{MixStyle}
MixStyle simulates new styles by mixing two instances' style statistics with a random convex weight sampled from the Beta distribution. First, this method shuffles or swaps the position of each sample in a batch to generate a reference batch $\mathcal{\tilde{B}}$ from the original batch $\mathcal{B}$. Then calculates the mixed feature statistics ($\gamma^{mix}$, $\beta^{mix}$) by 
\begin{equation}
    \begin{split}
        &\gamma^{mix}=\lambda\sigma(\textbf{x}_\mathcal{B}) + (1-\lambda)\sigma(\textbf{x}_\mathcal{\tilde{B}}), \\
        &\beta^{mix}=\lambda\mu(\textbf{x}_\mathcal{B}) + (1-\lambda)\mu(\textbf{x}_\mathcal{\tilde{B}}),
    \end{split}
\end{equation}
where $\lambda$ is also derived from the Beta distribution ($\lambda \sim Beta(\alpha,\alpha), \alpha\in(0,\infty)$), $\textbf{x}_\mathcal{B}$ is the instances in $\mathcal{\tilde{B}}$, and $\mu(\textbf{x})$ and $\sigma(\textbf{x})$ are the mean and standard deviation of $\textbf{x}$, respectively. Mixed statistics are applied to the style-normalized $\textbf{x}$ to construct a virtual data sample as 
\begin{equation}
    MixStyle(\mathcal{B})=\gamma^{mix} \frac{\textbf{x}-\mu(\textbf{x})}{\sigma(\textbf{x})} + \beta^{mix}.
\end{equation}
MixStyle is activated probabilistically and only changes instances' distribution, not the class labels.

\begin{table*}[t!]
    \caption{Overall drowsiness classification performance (\%) in subjects according to the augmentation methods}
    \centering
    \resizebox{2\columnwidth}{!}{%
    \begin{tabular}{lccccccccccc|c}
    \toprule
    \multirow{2}{*}{Model} & \multicolumn{11}{c|}{Subject} \\ \cmidrule{2-13}
     & S01 & S02 & S03 & S04 & S05 & S06 & S07 & S08 & S09 & S10 & S11 & Avg. \textit{F}1-score$\pm$Std. \\ \midrule
    ResNet1D-18 & 74.71 & 12.90 & 71.92 & 64.33 & 62.72 & 78.76 & \textbf{80.72} & \textbf{65.55} & 71.02 & 65.17 & 63.57 & 64.67$\pm$18.26 \\
    \hspace{4mm}+ Mixup \cite{zhang2018mixup} & \textbf{80.19} & 9.49 & \textbf{81.76} & 65.07 & \textbf{83.72} & 85.31 & 78.50 & 57.99 & 77.60 & 62.07 & 69.17 & 68.26$\pm$21.58 ($+$3.59)$^a$ \\
    \hspace{4mm}+ Manifold Mixup \cite{verma2019manifold} & 74.00 & 17.87 & 75.95 & 64.21 & 77.63 & 80.81 & 79.26 & 58.90 & \textbf{80.37} & 63.74 & \textbf{78.60} & 68.30$\pm$18.37 ($+$3.63) \\
    \hspace{4mm}+ MixStyle(123) \cite{zhou2021domain} & 71.74 & 11.84 & 75.80 & \textbf{66.67} & 78.34 & 80.81 & 80.18 & 62.37 & 71.60 & \textbf{70.59} & 71.10 & 67.37$\pm$19.25 ($+$2.70)\\
    \hspace{4mm}+ MixStyle(1234) \cite{kim2022dg} & 62.65 & \textbf{29.08} & 71.70 & 66.03 & 80.58 & \textbf{88.15} & 79.11 & 61.35 & 73.62 & 68.69 & 71.53 & \textbf{68.41$\pm$15.28 ($+$3.74)}\\ \bottomrule
    \multicolumn{13}{l}{Avg.: Average, Std.: Standard deviation}\\
    \multicolumn{13}{l}{MixStyle($\cdot$): The position where MixStyle is applied (e.g., MixStyle(123) defines applying MixStyle after the first, second, and third residual block)} \\
    \multicolumn{13}{l}{$^a$Numbers in parenthesis indicate the amount of improvement in \textit{F}1-score based on the baseline} \\
    \end{tabular}}  
    \label{tab:augment}
\end{table*}


\section{EXPERIMENTS}
\subsection{Dataset Description}
We used a publicly available dataset in \cite{cui2022eeg} that consists of eleven subjects' EEG signals. The subjects and sessions were selected from a dataset \cite{cao2019multi} conducted at the National Chiao Tung University, Taiwan. EEG signals were acquired while driving on an empty, straight road for 90 minutes. Participants had to focus on the road and steer the wheel when they recognized a lane deviation. A random lane-deviation event occurred five to ten seconds after the response offset. Each event moment, including the starting point of lane deviation, recognition point, and when the car reaches the original lane, is denoted as deviation onset, response onset, and response offset. In the original dataset, the direction of the lane-deviation event was additionally divided, but in the preprocessed dataset \cite{cui2022eeg}, they are considered the same event. 32 Ag/AgCl electrodes were used to record EEG signals at a sampling rate of 500 Hz.

\subsection{Data Preprocessing}
\begin{figure}
    \centering
    \includegraphics[width=0.83\columnwidth]{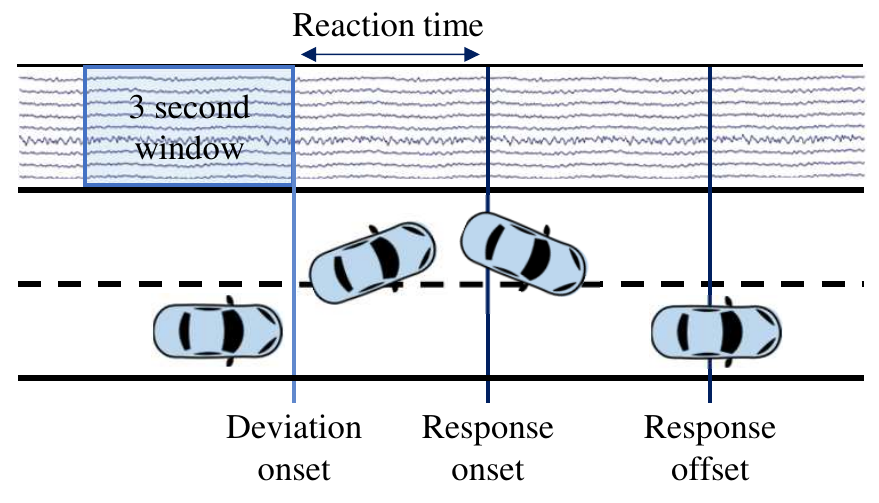}
    \caption{Experimental protocol of the publicly available dataset. A 3-second window before the lane deviation event was segmented. Deviation onset, response onset, and response offset denote the starting point of lane deviation, the recognition point, and the point when the original lane is reached.}
    \label{fig:dataset}
\end{figure}

We used the 'unbalanced dataset' \cite{cui2022eeg} that is down-sampled to 128 Hz and comprised of a 3-second window of data before the lane-deviation event, as in Fig. \ref{fig:dataset}. Data samples were labeled into two classes (`alert' and `drowsy') based on the reaction time (RT) \cite{wei2018toward, cui2022eeg}. RT is defined as the difference in time between the deviation onset and the response onset. The selected eleven subjects had a relatively balanced distribution of classes among the sessions, with each class comprising over 50 samples. In total, 1,221 and 1,731 samples for the `drowsy' class and the `alert' class, respectively.

\subsection{Implementation Details}
We evaluated the performance using leave-one-subject-out cross-validation \cite{kwon2019subject}. In detail, 90\% of each remaining subject data was used as a training set and 10\% as the validation set. At training, samples were drawn from all source domains in a manner that considered both domain and class balance \cite{gulrajani2021in} to consider the number of samples in each domain. We attempted to increase the influence of minority sample domains. As an optimizer, we used an Adam algorithm with a learning rate of 0.002. Furthermore, we assigned $\alpha=0.1$ for all of the Beta distributions used for augmentation and activated MixStyle with a probability of 50\%. In addition, we considered the `drowsy' class as positive and `alert' as negative for evaluation.

\begin{table}[t!]
    \caption{Average drowsiness classification performance (\%) of deep learning models}
    \centering
    \resizebox{0.85\columnwidth}{!}{%
    \begin{tabular}{lcccc}
    \toprule
    Model & \textit{F}1-score & AUROC & Precision & Recall \\ \midrule
    EEGNet4,2 \cite{lawhern2018eegnet} & 61.99 & 70.45 & 75.78 & 57.53 \\
    EEGNet8,2 \cite{lawhern2018eegnet} & 63.97 & \textbf{71.26} & 74.63 & 61.24 \\
    ResNet1D-8 \cite{han2021domain} & 64.12 & 68.83 & \textbf{71.10} & 63.40 \\
    ResNet1D-18 \cite{kim2022dg} & \textbf{64.67}& 69.33 & 70.25 & \textbf{64.49} \\ \bottomrule
    \end{tabular}}
    \label{tab:baseline}
\end{table}

\subsection{Results and Discussion}
As Table \ref{tab:baseline} shows the average drowsiness classification performance of deep learning models, ResNet1D-18 achieved the highest \textit{F}1-score of 64.67\% and AUROC of 69.33\%. Because \textit{F}1-score accounts more for positive classes, the higher \textit{F}1-score results mean that the model correctly classifies drivers' drowsiness as drowsy. Among EEGNet models, EEGNet8,2, which learns more temporal filters, achieved a higher \textit{F}1-score of 63.97\% and the highest AUROC of 71.26\%. Among ResNet1D models, the network with deeper layers achieved a higher generalization performance, an \textit{F}1-score of 64.67\% and AUROC of 69.33\%. 
As a result, we selected the ResNet1D-18 as our backbone network for comparing augmentation methods.

Table \ref{tab:augment} shows each subject's overall drowsiness classification performance according to the augmentation methods. The number between the parenthesis indicates the position where MixStyle is applied. For example, MixStyle(1234) defines using MixStyle after all the residual blocks. Comparing the average generalization performance, using MixStyle after every residual block, as in \cite{kim2022dg}, resulted in an \textit{F}1-score of 68.41\% with the smallest standard deviation of 15.28\%. The performance in each subject was not the best but showed competitive results. In addition, the average \textit{F}1-score increased by 3.59\%, 3.63\%, 2.70\%, and 3.74\% for applying Mixup, Manifold Mixup, and Mixstyle after the first three blocks and after all the residual blocks, respectively. Specifically, Mixup, a raw-level augmentation method, performed the lowest performance among methods of an \textit{F}1-score of 68.26\%, and MixStyle, a mainifold-level augmentation method, improved the overall highest generalization performance. Therefore, combining intermediate features for augmentation seems to be effective in EEG signal datasets. Moreover, comparing the placement of implementing MixStyle, unlike the computer vision tasks \cite{zhou2021domain} where the label-sensitive information is captured at the last block, style information remains until the last residual block in the publicly available dataset like \cite{kim2022dg}.

\section{CONCLUSION}
We proposed a robust calibration-free driver drowsiness classification framework by a manifold-level augmentation. Unseen domain samples were generated by mixing intermediate instances' style statistics. We conducted experiments with multiple augmentation methods, especially manifold-level augmentations. Based on the experiments, applying augmentation with intermediate features increased the framework's robustness. Furthermore, the position of augmentation influenced the improvement of EEG-based classification. Our limitations are that we compared only three augmentation algorithms, and this framework is limited to driver drowsiness classification. Therefore, for future analysis, we will benchmark other drivers' mental state-related datasets and compare other augmentation methods.

\bibliographystyle{IEEEtran}
\bibliography{REFERENCE}

\begin{thebibliography}{10}
\providecommand{\url}[1]{#1}
\csname url@samestyle\endcsname
\providecommand{\newblock}{\relax}
\providecommand{\bibinfo}[2]{#2}
\providecommand{\BIBentrySTDinterwordspacing}{\spaceskip=0pt\relax}
\providecommand{\BIBentryALTinterwordstretchfactor}{4}
\providecommand{\BIBentryALTinterwordspacing}{\spaceskip=\fontdimen2\font plus
\BIBentryALTinterwordstretchfactor\fontdimen3\font minus
  \fontdimen4\font\relax}
\providecommand{\BIBforeignlanguage}[2]{{%
\expandafter\ifx\csname l@#1\endcsname\relax
\typeout{** WARNING: IEEEtran.bst: No hyphenation pattern has been}%
\typeout{** loaded for the language `#1'. Using the pattern for}%
\typeout{** the default language instead.}%
\else
\language=\csname l@#1\endcsname
\fi
#2}}
\providecommand{\BIBdecl}{\relax}
\BIBdecl

\bibitem{won2017motion}
D.-O. Won, H.-J. Hwang, D.-M. Kim, K.-R. M{\"u}ller, and S.-W. Lee,
  ``Motion-based rapid serial visual presentation for gaze-independent
  brain-computer interfaces,'' \emph{IEEE Trans. Neural Syst. Rehabil. Eng.},
  vol.~26, no.~2, pp. 334--343, 2017.

\bibitem{gao2019eeg}
Z.~Gao \emph{et~al.}, ``{EEG}-based spatio-temporal convolutional neural
  network for driver fatigue evaluation,'' \emph{IEEE Trans. Neural Netw.
  Learn. Syst.}, vol.~30, no.~9, pp. 2755--2763, 2019.

\bibitem{lee2019connectivity}
M.~Lee \emph{et~al.}, ``Connectivity differences between consciousness and
  unconsciousness in non-rapid eye movement sleep: {a TMS-EEG} study,''
  \emph{Sci. Rep.}, vol.~9, no.~1, pp. 1--9, 2019.

\bibitem{wu2020transfer}
D.~Wu, Y.~Xu, and B.-L. Lu, ``Transfer learning for {EEG}-based brain-computer
  interfaces: A review of progress made since 2016,'' \emph{IEEE Trans. Cogn.
  Dev. Syst.}, 2020.

\bibitem{thung2018conversion}
K.-H. Thung \emph{et~al.}, ``Conversion and time-to-conversion predictions of
  mild cognitive impairment using low-rank affinity pursuit denoising and
  matrix completion,'' \emph{Med. Image Anal.}, vol.~45, pp. 68--82, 2018.

\bibitem{zhang2017hybrid}
Y.~Zhang, H.~Zhang, X.~Chen, S.-W. Lee, and D.~Shen, ``Hybrid high-order
  functional connectivity networks using resting-state functional {MRI} for
  mild cognitive impairment diagnosis,'' \emph{Sci. Rep.}, vol.~7, no.~1, pp.
  1--15, 2017.

\bibitem{xu2021key}
T.~Xu \emph{et~al.}, ``E-key: {An} {EEG}-based biometric authentication and
  driving fatigue detection system,'' \emph{IEEE Trans. Affect. Comput.}, 2021.

\bibitem{paulo2021cross}
J.~R. Paulo, G.~Pires, and U.~J. Nunes, ``Cross-subject zero calibration
  driver’s drowsiness detection: {Exploring} spatiotemporal image encoding of
  {EEG} signals for convolutional neural network classification,'' \emph{IEEE
  Trans. Neural Syst. Rehabil. Eng.}, vol.~29, pp. 905--915, 2021.

\bibitem{kwon2019subject}
O.-Y. Kwon, M.-H. Lee, C.~Guan, and S.-W. Lee, ``Subject-independent
  brain-computer interfaces based on deep convolutional neural networks,''
  \emph{IEEE Trans. Neural Netw. Learn. Syst.}, vol.~31, no.~10, pp.
  3839--3852, 2019.

\bibitem{cui2019eeg}
Y.~Cui, Y.~Xu, and D.~Wu, ``{EEG}-based driver drowsiness estimation using
  feature weighted episodic training,'' \emph{IEEE Trans. Neural Syst. Rehabil.
  Eng.}, vol.~27, no.~11, pp. 2263--2273, 2019.

\bibitem{kim2019subject}
K.-T. Kim, C.~Guan, and S.-W. Lee, ``A subject-transfer framework based on
  single-trial {EMG} analysis using convolutional neural networks,'' \emph{IEEE
  Trans. Neural Netw. Learn. Syst.}, vol.~28, no.~1, pp. 94--103, 2019.

\bibitem{lee2018high}
M.-H. Lee, J.~Williamson, D.-O. Won, S.~Fazli, and S.-W. Lee, ``A high
  performance spelling system based on {EEG-EOG} signals with visual
  feedback,'' \emph{IEEE Trans. Neural Syst. Rehabil. Eng.}, vol.~26, no.~7,
  pp. 1443--1459, 2018.

\bibitem{suk2014predicting}
H.-I. Suk \emph{et~al.}, ``Predicting {BCI} subject performance using
  probabilistic spatio-temporal filters,'' \emph{PloS one}, vol.~9, no.~2, p.
  e87056, 2014.

\bibitem{kostas2020thinker}
D.~Kostas and F.~Rudzicz, ``Thinker invariance: Enabling deep neural networks
  for {BCI} across more people,'' \emph{J. Neural Eng.}, vol.~17, no.~5, p.
  056008, 2020.

\bibitem{liu2020inter}
Y.~Liu, Z.~Lan, J.~Cui, O.~Sourina, and W.~M{\"u}ller-Wittig, ``Inter-subject
  transfer learning for {EEG}-based mental fatigue recognition,'' \emph{Adv.
  Eng. Inform.}, vol.~46, p. 101157, 2020.

\bibitem{han2021domain}
D.-K. Han and J.-H. Jeong, ``Domain generalization for session-independent
  brain-computer interface,'' in \emph{Int. Winter Conf. Brain-Computer
  Interface (BCI)}, 2021, pp. 1--5.

\bibitem{hwang2021mitigating}
S.~Hwang, S.~Park, D.~Kim, J.~Lee, and H.~Byun, ``Mitigating inter-subject
  brain signal variability for {EEG}-based driver fatigue state
  classification,'' in \emph{IEEE Int. Conf. Acoust. Speech, Signal Process.
  (ICASSP)}, Toronto, Ontario, Canada, June 2021, pp. 990--994.

\bibitem{gulrajani2021in}
I.~Gulrajani and D.~Lopez-Paz, ``In search of lost domain generalization,'' in
  \emph{Int. Conf. Learn. Represent. (ICLR)}, 2021.

\bibitem{dou2019domain}
Q.~Dou, D.~C. de~Castro, K.~Kamnitsas, and B.~Glocker, ``Domain generalization
  via model-agnostic learning of semantic features,'' in \emph{Proc. Adv.
  Neural Inf. Process. Syst. (NeurIPS)}, vol.~32, 2019.

\bibitem{zhou2022domain}
K.~Zhou, Z.~Liu, Y.~Qiao, T.~Xiang, and C.~C. Loy, ``Domain generalization: A
  survey,'' \emph{IEEE Trans. Pattern Anal. Mach. Intell.}, 2022.

\bibitem{lawhern2018eegnet}
V.~J. Lawhern \emph{et~al.}, ``{EEGNet: A} compact convolutional neural network
  for {EEG}-based brain-computer interfaces,'' \emph{J. Neural Eng.}, vol.~15,
  no.~5, p. 056013, 2018.

\bibitem{cheng2020subject}
J.~Y. Cheng, H.~Goh, K.~Dogrusoz, O.~Tuzel, and E.~Azemi, ``Subject-aware
  contrastive learning for biosignals,'' \emph{arXiv preprint
  arXiv:2007.04871}, 2020.

\bibitem{jeong2020brain}
J.-H. Jeong, K.-H. Shim, D.-J. Kim, and S.-W. Lee, ``Brain-controlled robotic
  arm system based on multi-directional {CNN-BiLSTM} network using {EEG}
  signals,'' \emph{IEEE Trans. Neural Syst. Rehabil. Eng.}, vol.~28, no.~5, pp.
  1226--1238, 2020.

\bibitem{jeong2020decoding}
J.-H. Jeong, N.-S. Kwak, C.~Guan, and S.-W. Lee, ``Decoding movement-related
  cortical potentials based on subject-dependent and section-wise spectral
  filtering,'' \emph{IEEE Trans. Neural Syst. Rehabil. Eng.}, vol.~28, no.~3,
  pp. 687--698, 2020.

\bibitem{he2016deep}
K.~He, X.~Zhang, S.~Ren, and J.~Sun, ``Deep residual learning for image
  recognition,'' in \emph{Proc. IEEE Conf. Comput. Vis. Pattern Recognit.
  (CVPR)}, 2016, pp. 770--778.

\bibitem{kim2022dg}
D.-Y. Kim, D.-K. Han, J.-H. Jeong, and S.-W. Lee, ``{EEG}-based driver
  drowsiness classification via calibration-free framework with domain
  generalization,'' in \emph{Proc. IEEE Int. Conf. Syst. Man, Cybern. (SMC)},
  2022, pp. 2293--2298.

\bibitem{zhang2018mixup}
H.~Zhang, M.~Cisse, Y.~N. Dauphin, and D.~Lopez-Paz, ``Mixup: {Beyond}
  empirical risk minimization,'' in \emph{Proc. Int. Conf. Learn. Represent.
  (ICLR)}, 2018.

\bibitem{verma2019manifold}
V.~Verma \emph{et~al.}, ``Manifold mixup: {Better} representations by
  interpolating hidden states,'' in \emph{Proc. Int. Conf. Mach. Learn.
  (ICML)}, 2019, pp. 6438--6447.

\bibitem{zhou2021domain}
K.~Zhou, Y.~Yang, Y.~Qiao, and T.~Xiang, ``Domain generalization with
  mixstyle,'' in \emph{Proc. Int. Conf. Learn. Represent. (ICLR)}, 2021.

\bibitem{cui2022eeg}
J.~Cui, Z.~Lan, O.~Sourina, and W.~M{\"u}ller-Wittig, ``{EEG}-based
  cross-subject driver drowsiness recognition with an interpretable
  convolutional neural network,'' \emph{IEEE Trans. Neural Netw. Learn. Syst.},
  2022.

\bibitem{cao2019multi}
Z.~Cao, C.-H. Chuang, J.-K. King, and C.-T. Lin, ``Multi-channel {EEG}
  recordings during a sustained-attention driving task,'' \emph{Sci. Data},
  vol.~6, no.~1, pp. 1--8, 2019.

\bibitem{wei2018toward}
C.-S. Wei, Y.-T. Wang, C.-T. Lin, and T.-P. Jung, ``Toward drowsiness detection
  using non-hair-bearing {EEG}-based brain-computer interfaces,'' \emph{IEEE
  Trans. Neural Syst. Rehabil. Eng.}, vol.~26, no.~2, pp. 400--406, 2018.

\end{thebibliography}

\end{document}